# Natural Economics

## Some Different Ideas

Louis Mello

## Introduction

Infused into the lives of billions of distinct denizens of this planet is the practice of economic affairs, based primarily on the need to satisfy essential elements of survival and on the need to assure the continuation of the species. This is represented by a mathematically definable branching process underscored by countless daily *transactions* which, when taken separately, appear to be virtually meaningless (or random), however, when seen as part of a much larger system, take on the characteristics of a *complex organism*.

In order to understand this *organism*, that society has convened to call Economics, one must search for its driving force, one must comprehend the minutiae as well as the broader concepts while concurrently baring the beast in its most primitive state.

The vantage point whence one must purview this matter is, to paraphrase Heisenberg, from where one does not interfere with what is observed. Thus, it is useless to attempt to impart any measure of judgment on the system.

Regard the figure below: an irregular curve that winds its way through time in fractured form.

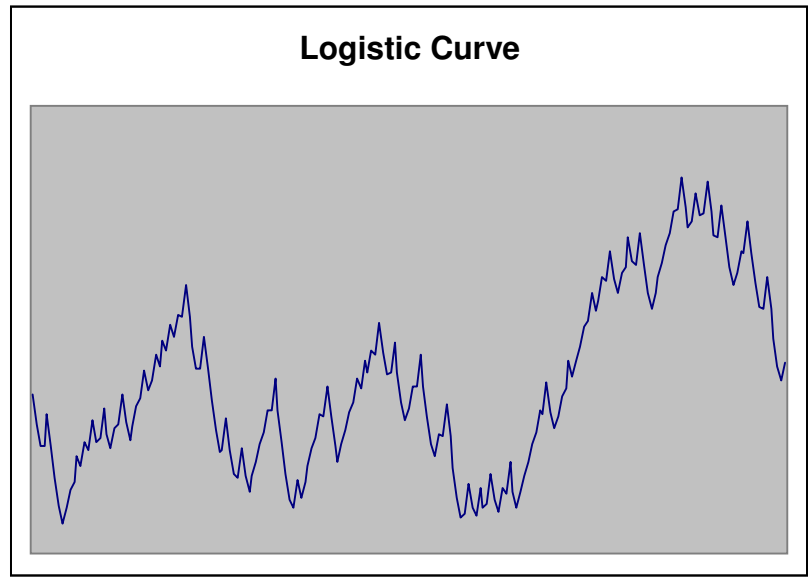

**Logistic Curve**

This is history, in discrete units of time, feeding on its own output. Essentially taking *reality* in $t$-1 and using it as input in $t$. In other words, it is the immediate memory of what has just occurred influencing what is just about to happen.

Initial visual inspection yields the familiar notion that there are periods of growth or prosperity followed by periods of recession; *irregular dynamic cycles* that define the behavior of the curve itself. But so that we may better understand these *cycles* we must espouse the Archimedean view – a perspective that is not, in any way, associated with the run-of-the-mill anthropocentric idealization of our surroundings. As such, one must seek out *the* pattern that would persist independently of the observer, and by definition, there must only be one.

Thus, this curve, when stripped of the effects of human hermeneutics, displays a splendorous wealth of detail – a self-similarity – that is so often found in nature. This nonlinearity is at the core of all things, man-made or not. The complexity may at first seem overwhelming, yet its origin is mathematically quite simple:

$$x_{t+1} = x_t r \left(1 - x_t\right) \quad (1.1)$$

This metamorphosis from simple to complex can be shown to occur without the presence of exogenous forces. *This means, in the realm of economic organization, that systems exist in and of themselves – human players are mere accidents in their unfolding. They are, as we shall see further along, Self-Organizing Systems.*

Bold as this statement may seem, bolder still is to acknowledge the daunting reality that *control* is an illusion: the unbounded hours spent devising "new and improved" formulae aimed at domesticating such unruly conduct are no more useful than attempting to reach the moon in a sailboat. Indeed, all such travail only exposes more blatantly the subject's contamination by the observer.

Economics, as it is practiced in our day, is nothing more than exacerbated social grammar, spawned by monitoring convention and association *post factum* and then proceeding to derive "rules" that *seem* to fit the data. And given that what one sees is so evidently nonlinear, and nonlinearity is so damnably difficult to tame with prevailing mathematical tools, economists – not unlike conjurers and alchemists of old – employ linear models in the vain hope that the nonlinear character of the system will dissipate in the manner of clouds after a summer squall. But it does not fit; it cannot be linearized to tally to the world as seen through human eyes. In no other field of study is it more important to *take things as they are* – raw and unpolished – and examine them under the prism of current nonlinear theory.

Economics is not comparable to physics; it is less structured, more fluid, and much fuzzier. However, it is part of nature, and most certainly should be respected, maybe even feared, at the risk of consequences that could lead to the extinction of a species.

## Free-Will
Nothing could be farther from what is commonly called "rational" than human behavior. Economics has steadied itself on the shoulders of *Homo Economicus Rationalis* with quasi dogmatic fervor and the results have been abysmal.

> *"For my own part I am now somewhat skeptical of the success of a merely monetary policy directed towards influencing the interest rate. I expect to see the State, which is in a position to calculate the marginal efficiency of capital-goods on long views and on the basis of the general social advantage, taking an ever greater responsibility for directly organizing investment; since it seems likely that the fluctuations in the market estimation of the marginal efficiency of different types of capital, calculated on the principles that I have described above, will be too great to be offset by any practicable changes in the rate of interest."*[1]

Even the most forward-looking scientists, men such as Ilya Prigogine and Richard Day, have stumbled on this gulf that separates volition from absolute determinism.

> *"The point to be emphasized here is that economic phenomena introduce volition in a fundamental way. The atoms (people), governed by volition, form molecules and compounds (households, firms, government agencies) according to mentally emergent laws that evolve: laws governed by legal or quasi legal processes with new institutions being invented and with existing ones undergoing continual revision and transformation.*
> 
> *All what we want to say is that the operation of mind governs the flow of material in the economic world. It is as if atoms could decide what kind of atom they want to be: what valence, what atomic weight; could decide with whom they want to form molecules; could invent completely new molecules that never existed before; or could set off motion into orbits of their choosing and return to some other trajectory at will. Those who hope to discover laws of human behavior will have to deal with the dynamics of organizations and the mechanisms that lead people to obey or violate the laws they create for themselves."*[2]

The nucleus of the matter is, doubtless, *free will*. Does one really have a choice in the process of decision-making? Are the multiple transactions, billions every second of every day, the product of some choice predicated upon the concept that it represents what is best for any one group of people?

Science and history – based on empirical evidence – retort with a resounding NO! *Free will*, on the economic playing field, means that choices must be dictated by variables such as income, geographies and an inestimable host of nameless factors that determine what is available and what is affordable. If this were not so marketing executives would not keep so close a watch on demographics; the futures markets would cease to be; all form of speculation would yield no fruit. The economy, as it is presented by the wealthiest 5%, would not be able to function, simply because there would be no one to buy that which is in their best interest to sell.

To reconcile *free will* and determinism one must reject the idea of exogenous shocks to the system and consider only endogenous feedback as its fuel; in much the same way that our equation (1.1) demonstrates. External pressure is unnecessary – determinism is engendered by the inner workings of the system alone or, in this case, the interaction of the particles (individuals) as they are driven by the information currently available.

---

[1] Keynes, John Maynard *The General Theory if Employment, Interest and Money* (1935)
[2] Richard Day, *Physics and the Foundations of Economic Science*, 2003

*Free will* is, alongside randomness, the outward appearance of choice brought on by an overall lack of information/knowledge of the system, especially with respect to long term expectation.

In the years preceding the depression of 1929 did some variety of mass delusion occlude the American mind? Irrational as the spiral of unchecked credit mongering was, **no one believed it would implode** leaving utter devastation in its path. No one thought that prices on the Stock Market would ever go down; but they did.

Reaching back farther into the past, those who sold entire estates to purchase a single tulip bulb believed, if nothing else, that they would sell the very same bulb for more.

Are these the actions of "rational" beings? Even more; did the crazed citizenry depicted in these bubbles truly have a choice?

Indeed they did not; in every instance of boom-and-bust to which the somewhat unsuspecting earth-dweller has been subjected, there has been a **not so invisible hand** guiding him toward the proverbial pot of gold. Thus, the perceived irrationality, in fact, is described by the lure of facile riches that lines the very spirit of the system. This is the only way to observe the system without measurement interfering in its trajectory.

## Self-Organizing Systems

The essence of self-organization is that system[3] structure (at least in part) appears with no explicit pressure or constraints from outside itself; the constraints on form are internal to the system and result from the interactions between the components, whilst being independent of their physical nature. The organization can evolve in either time or space, it can maintain a stable form or, it can demonstrate transient phenomena. General resource flows into or out of the system are permitted, but are not critical to the concept.

*Self-organization, thus, seeks to discover the general rules under which such structures appear;* the forms under which they function and the methods of predicting change in structure resulting from change in the fundamental system. The results are expected to be applicable to any system exhibiting the same network characteristics.

---

[3] A system is a collection of interacting parts performing as a whole. It can be distinguished from its surroundings because of recognizable boundaries. The function depends on the arrangement of the parts and will change in some way if parts are added, removed or rearranged. Such systems have properties that are emergent, i.e., not contained within any of the parts. They exist only at a higher level of description.

As is the case with any self-organizing system, the effects of exogenous interference are, at best, limited in scope and efficacy. But, does this mean that any and all external meddling or tweaking is utterly innocuous? No, it does not. Consider traffic engineering: arguably a necessary evil in every city with more than 20 vehicles. While any manner of signals or signs can serve as an aid to drivers, in the final analysis the individual motorists will do what is in their best interest, including breaching traffic laws when appropriate. However they will respect the traffic guidelines inasmuch as they contribute to their plan, which is to get from point A to point B without injuring themselves or others. The practice of economic interaction is, similarly, guided by principles that will assist the distinct consumer in the satisfaction of specific needs. It is, thus, exceptionally clear that only when economic systems are controlled by any one group (as could be the case of, say, truck drivers in the traffic analogy or large economic entities such as Exxon or Halliburton, in our present analysis) is there a decided disavowal of Free Will, hence, a fundamental malfunction in the *governing dynamics*[4] of the system. People, when left to their own devices, will find a way to trade, barter, buy, sell, produce, consume and otherwise procure economic satisfaction. Primitive human communities are living proof of that.

This, then, leads us to one of the most damaging aspects of the current capitalist economic organization: the fact that it denies *emergence*, i.e. the appearance of a property[5] or feature of the system not previously observed. Generally, higher level properties are regarded as emergent -- a car is an emergent property of the interconnected parts. That property disappears if the parts are disassembled and just placed in a heap. This feature of self-organizing systems is what allows new forms of relationship between the parts to *emerge* in an evolutionary mode. It also clears the path for the dialectical correlation between elements of a system to transform said elements and then, dynamically have them reappear as a "new" set of rules.

The mathematical construct for such behavior is embodied in the notion of an attractor. This is a preferred position for the system, such that, if the system is started from another state, it will evolve until it arrives at the attractor, and will then stay there in the absence of other factors. An attractor can be a point (e.g. the center of a bowl containing a ball), a regular path (e.g. a planetary orbit), a complex series of states (e.g. the metabolism of a cell) or an infinite sequence (called a *strange attractor*). All of these are exemplars of a restricted volume of state space. The area of state space that leads to an attractor is called its basin of attraction.

---

[4] A mathematical concept developed by John Nash.

[5] If we connect a series of parts in a loop, then that loop does not exist as a property of the parts themselves. The parts can have any structure or form and yet the loop persists. If the loop confirms additional dynamic behavior (it might oscillate, for example) then it is an instance of an emergent system property.

The analogy is obvious: the state space is a set of socio-economic rules generally accepted by the members of society at large. The four different types of attractor correspond to different phases in economic development or, different systems, as long as we admit that economics is in a state of process and not of being. Notably, the *strange attractor* is usually more perceptible near a phase transition, i.e., a point in time when the system in undergoing large-scale transformation.

Any system that moves to a fixed structure can be said to be drawn to an attractor. In our case, when the economic system is dominated by some exceptionally strong force, we observe the existence of a fixed point attractor. A complex system should have many attractors and these may alter with transformations to the system's interconnections (mutations). Certainly, modern economic interaction qualifies as such a complex system, and so we study self-organization in order to investigate the attractors of the system, their form and dynamics. It is plain, thus, that the fixed point attractor is an unnatural state for a complex system.

The example presented in the Introduction (Eq. 1.1) is a simple *strange attractor*. It illustrates how complexity can arise from very basic rules. The observed dynamics of all economic relationships fit the description offered by complexity better than any other. It is only the straight-jacket that conservative (static) economists attempt to use to control the economy that is *contra natura*. This is emphatically not, by any imaginable stretch of the imagination, an apology of *laissez-faire* capitalism. History has proven, beyond all doubt, that the Adam Smith's invisible hand is not invisible at all.

What is sought here is a framework to rethink the logical tenets that currently constrain economic thought; a more thorough understanding of the role of the individual in this process and most significantly, a more realistic (natural) approach to the laws that govern the dynamics of economic behavior.